\documentstyle[11pt,aaspp4]{article}

\def \Mo {\ifmmode M_\odot \else $M_\odot$ \fi}
\def \Ro {\ifmmode R_\odot \else $R_\odot$ \fi}
\def \Lo {\ifmmode L_\odot \else $L_\odot$ \fi}
\def \Re {\ifmmode R_{\sc e} \else $R_{\sc e}$ \fi}
\def \Oe {\ifmmode \theta_{\sc e} \else $\theta_{\sc e}$ \fi}
\def \Rs {\ifmmode R_{\sc s} \else $R_{\sc s}$ \fi}
\def \SgrA {\ifmmode {\rm SgrA}^\star\, \else SgrA$^\star$\fi}
\def \avm {\bar{m}}
\def \avL {\bar{L}_K}
\def \avv {\bar{v}_2}

\def \avu {\langle u_0 \rangle}
\def \arcsec {^{\prime\prime}}
\def \MLKo {\Upsilon_\odot}
\def \mag {^{\rm m}}
\def \DT {\Delta T}
\def \DK {\Delta K}
\def \Dp {\Delta p}
\def \nuk {\nu}
\def \LK {L_K}
\def \La {L_l}
\def \Lb {L_u}
\def \Lt {L_0}
	
\def \tvt {\hat{v}_2}
\def \tvmax {\hat{v}_{\rm max}}
\def \tvrot {\hat{v}_{\rm rot}}

\lefthead{Alexander and Sternberg}
\righthead{Near-Infrared Microlensing by the Black Hole in the Galactic Center}

\begin{document}
\title{Near-Infrared Microlensing of Stars by the Super-Massive
Black Hole in the Galactic Center}
\author{Tal Alexander\altaffilmark{1}}
\affil{Institute for Advanced Study, Olden Lane, Princeton, NJ 08540}
\and
\author{Amiel Sternberg\altaffilmark{2,3}}
\affil{School of Physics and Astronomy, Tel Aviv University, Ramat Aviv, 
Tel Aviv 69978, Israel}
\altaffiltext{1}{E-mail address: tal@ias.edu}
\altaffiltext{2}{E-mail address: amiel@wise.tau.ac.il}
\altaffiltext{3}{On leave at the Department of Astronomy, University
of California, Berkeley, CA 94720}
\authoremail{tal@ias.edu}
\begin{abstract} 
We investigate microlensing amplification of faint stars in the dense
stellar cluster in the Galactic Center by the super-massive black hole,
which is thought to coincide with the radio source $\SgrA$. Such
amplification events would appear very close to the position of $\SgrA$,
and could be observed, in principle, during the monitoring of stellar
proper motions in the Galactic Center.

We use observations of the near-infrared $K$-band (2.2\,$\mu$m) luminosity
function in the Galactic Center and in Baade's Window, as well as stellar
population synthesis computations, to construct empirical and theoretical
$K$ luminosity function models for the inner 300\,pc of the Galaxy. These,
together with the observed dynamical properties of the central cluster and
inner bulge, are used to compute the rates of microlensing events which
amplify stars with different intrinsic luminosities above specified
detection thresholds.

We present computations of the lensing rates as functions of the event
durations, which range from several weeks to a few years, for detection
thresholds ranging from $K_0=16\mag$ to $19\mag$. We find that events with
durations shorter than a few months dominate the lensing rate because of
the very high stellar densities and velocities very near the black hole,
where the effective lens size is small. For the current detection limit of
$K_0=17\mag$, the total microlensing rate is $3\times 10^{-3}$ yr$^{-1}$.
The rate of events with durations $\ge 1$ yr is $1\times 10^{-3}$
yr$^{-1}$. The median value of the peak amplification for short events is
$\DK \sim 0.75\mag$ above the detection threshold, and is only weakly
dependent on $K_0$.  Long events are rarer, and are associated with more
distant stars, stars at the low velocity tail of the velocity distribution,
or stars that cross closer to the line-of-sight to $\SgrA$. Therefore, the
median peak amplifications of long events are larger and attain values $\DK
\sim 1.5\mag$ above the threshold.

Recent proper motion studies of stars in the Galactic Center have revealed
the possible presence of one or two variable $K$-band sources very close
to, or coincident with, the position of $\SgrA$ (Eckart \& Genzel
1996; Genzel et al. 1997; Ghez et al. 1998).  These sources may have
attained peak brightnesses of $K\approx15\mag$, about $1.5$--$2\mag$ above
the observational detection limits, and appear to have varied on a
timescale of $\sim 1$ yr. This behavior is consistent with long-duration
microlensing amplification of faint stars by the central black
hole. However, we estimate that the probability that a single such an event
could have been detected during the course of the recent proper motion
monitoring campaigns is $\sim 0.5\%$.  A ten-fold improvement in the
detection limit and $10\,$yr of monthly monitoring could increase the total
detection probability to $\sim20\%$.

\end{abstract}
\section{Introduction}

Recent proper motion studies of infrared-luminous stars in the Galactic
Center (GC) (\cite{EG96,Genzel97,Ghez98}) have convincingly demonstrated the
existence of a compact $\sim2.6\,10^6\Mo$ dark mass in the dynamical center
of the GC, which is located within $0.1\arcsec$ of the radio source $\SgrA$
(\cite{Ghez98}) ($1\arcsec=0.039\,$pc at 8\,kpc). Lower bounds on the
compactness of this mass concentration, together with dynamical
considerations, argue against the possibility of a massive cluster, and
point towards a super-massive black hole as the likely alternative
(\cite{Genzel97}; \cite{Maoz98}). This conclusion has important
implications for basic issues such as the prevalence of massive black holes
in the nuclei of normal galaxies, and the nature of the accretion mechanism
that makes $\SgrA$ so much fainter than typical active galactic nuclei
(\cite{Melia94}; \cite{NYM95}). 


Wardle \& Yusef-Zadeh (1992) considered several gravitational lensing
effects that could be induced by a massive black hole in the GC.
In particular, Wardle \& Yusef-Zadeh pointed out that gravitational lensing
would occasionally lead to the amplification and splitting of the stellar
images of stars which happen to move behind the black hole, and that such
events would typically last from months to years, depending on the distance
of such stars from the black hole. Wardle \& Yusef-Zadeh also estimated
that an optical depth of order unity for such events requires an observed
central surface density $>1000$ stars arcsec$^{-2}$, and that this in turn
would require an angular resolution of $\la 0.01\arcsec$ to individually
separate the lensed images from the crowded background of faint stars.
However, the photometric sensitivities and spatial resolutions required for
such observations are far beyond current observational
capabilities. Presently, the deepest near-infrared $K$-band (2.2 $\mu$m)
images of the central arcsec reach down to $K=17\mag$ (\cite{Ghez98}).  As
we will show, at this limiting magnitude the expected central surface
density is $\sim 20$ stars arcsec$^{-2}$ (\cite{Davidge97}).  The highest
spatial resolutions obtained so far are the diffraction limited resolutions
of $\sim 0.15\arcsec$ (\cite{Eckart95}; \cite{Davidge97}) and $\sim
0.05\arcsec$ (\cite{Ghez98}) achieved in the proper motion surveys.

In this paper we investigate a different possibility, namely {\it
microlensing} amplification of faint stars by the central black hole.  Such
events occur when the amplified but {\it unresolved} images of faint stars
rise above the detection threshold and then fade again as such stars move
behind the black hole close to the line-of-sight to $\SgrA$. We present a
quantitative study of such microlensing events, and we compute in detail
the microlensing rates as functions of the event durations, for a wide
range of assumed detection thresholds. In addition, we also consider the
possible amplification of sources which lie above the detection thresholds,
and we re-examine the limit in which the two lensed images can actually be
resolved.

Our study is motivated by several recent developments. Deep infrared star
counts in the inner GC (\cite{Blum96}; \cite{Davidge97}) and infrared and
optical star counts in Baade's Window (\cite{Tiede95}; \cite{Holtzman98})
now make it possible to reliably model the infrared stellar luminosity
function in the vicinity of the Galactic Center.  The on-going proper
motion monitoring campaigns of the inner few arcsec of the GC record both
the positions and fluxes of individual stars in the field at a sampling
rate of 1--2 observing runs per year. As a by-product, these measurements
can be used to search for microlensing events, albeit at a low sampling
rate. Such events would appear as time varying sources very close to
$\SgrA$.  It is therefore intriguing that one or two variable IR sources
may have already been detected close to, or coincident with, the position
of $\SgrA$ (\cite{Genzel97}; \cite{Ghez98}).  These sources may have
brightened to $K\sim15\mag$ before fading from view, and appear to have
varied on a timescale of $\sim 1$ yr.  As we will argue, this behavior is
consistent with the expected behavior of bright long-duration microlensing
events. However, we will also argue that the probability that a single such
event could have been detected during the course of the recent
proper-motion monitoring campaigns is small ($\sim 0.5\%$).  The detection
probabilities will increase considerably as the observational sensitivities
improve (e.g. with the advent of adaptive optics), and if $\SgrA$ is
monitored more frequently than has been done so far.

The structure of our paper is as follows. In \S\ref{s:rate} we set up the
formalism required for our computations. Some of the more technical aspects
are discussed in the appendix.  In \S\ref{s:gc} we discuss our treatment of
the stellar densities, velocity fields, and $K$-band luminosity functions,
which enter into the computation of the lensing rates. We present our
results in \S\ref{s:results}, where we provide computations of the lensing
rates as functions of the event durations for a wide range of assumed
detection thresholds. In \S\ref{s:discuss} we present a discussion and
summary.

\section{Lensing by the super-massive black hole} 
\label{s:rate}

The effective size of a gravitational lens at the lens plane is set by
the Einstein radius, $\Re$,
\begin{equation}
\Re = \left(\frac{4GM_\bullet}{c^2}\frac{Dd}{D+d}\right)^{1/2}
\sim 2.2\,10^{15} (M_{2.6} d_1)^{1/2}\,{\rm cm}\,,
\label{e:Re}
\end{equation}
where $G$ is the gravitational constant, $c$ is the speed of light,
$M_\bullet$ is the lens mass (here the black hole mass), $D$ and $d$ are
the observer--lens and lens--source distances, respectively (see
e.g. review by Bartelmann \& Narayan 1998). We assume, as will be justified
below, that $d
\ll D$, and define $M_{2.6} = M_\bullet/2.6\,10^6\Mo$ and $d_1 = d/1\,$pc. The
effective size of the lens at the source plane is $\Rs =
\Re(D+d)/D\sim\Re$. The angular size of the Einstein radius, $\Oe$, is
\begin{equation}
\Oe \sim 0.018\arcsec D_8^{-1}(M_{2.6}d_1)^{1/2}\,,
\label{e:Oe}
\end{equation}
where $8D_8\,{\rm kpc}$ is the Sun's galactocentric distance
(\cite{Carney95}).

In our study we assume that any IR-extinction associated with an accretion
disk or an `atmosphere' near the event horizon is negligible.

We begin by showing that in our problem the stars can be treated as point
sources and that the linear approximation (small light-bending angle
approximation) holds. First, when $d\ll D$, a star can be approximated as a
point source as long as its radius, $R_\star$, is much smaller than
$\Re/A$, where $A$ is the required amplification to observe the lensed
source above the detection threshold (see \S\ref{s:mlrate}). As we show
below (\S\ref{s:plklf}), the faintest stars that contribute significantly
to the lensing are low-mass stars with $R_\star \la 10^{11}$\,cm, which
require amplification factors of $A \la 100$.  Brighter stars require
progressively smaller amplification factors, down to $A\sim 1$, which for
detection thresholds $K\sim 17\mag$ correspond to stars with $R_\star < 
10^{12}$\,cm. Assuming a mean stellar mass of $\sim1\Mo$, a
central mass density of $\rho\sim4\,10^6\,\Mo$pc$^{-3}$ in the GC
(\cite{Genzel96}) implies a mean stellar separation of $\delta\sim
0.005\,$pc. Even at such a small distance from the black hole, $\Re\sim
1.5\,10^{14}\,$cm, so that the condition $R_\star\ll\Re/A$ is generally
satisfied for all relevant stars in the system. We thus conclude that the
point source approximation holds over the entire range of interest.

Second, the linear approximation holds as long as the Einstein radius is
much larger than the Schwarzschild radius of the black hole, $R_\bullet$,
\begin{equation}
\Re/R_\bullet \sim 
c\left(d/GM\right)^{1/2} \sim 2.8\,10^3(d_1/M_{2.6})^{1/2}\,,
\label{e:Rsch}
\end{equation}
which even for $d=\delta$ is as high as $\sim200$.

A point lens produces two images, one within and one outside the
Einstein radius.
The angular separation between the two images is
\begin{equation}
\Delta\theta = \Oe\sqrt{u^2+4}\,,
\label{e:Dtheta}
\end{equation}
and their mean angular offset from the optical axis is $u\Oe$, where $u\Oe$
is the angular separation between the unlensed source and the lens. As will
be shown in \S\ref{s:gc}, $u\ll1$ for amplification above present-day
detection thresholds in the GC, so that $\Delta\theta \sim 2\Oe$.

Three angular scales in the problem determine the way the lensing will
appear to the observer: the Einstein angle $\Oe(d)$, the FWHM angular
resolution of the observations, $\phi$, and the mean projected angular
separation between the observed stars, $\Dp(K_0)$, where $K_0$ is the
detection threshold. 

The lensed images have to be detected against the background of a very
dense stellar system. This background will be low as long as as the central
surface density of observed stars, $S_0 =\Dp^{-2}$, is small enough so
that $\pi \phi^2 S_0 < 1$. Thus, for a given detection threshold, an
angular resolution of at least
\begin{equation} 
\phi = \Dp(K_0)/\sqrt{\pi}\,.
\label{e:phi}
\end{equation} 
is required to detect the lensed star.  Lower resolutions correspond to the
regime of `pixel lensing' (\cite{Crotts92}), which we do not consider here.

When $\Oe < \phi$, the two images will not be resolved\footnote{We are
assuming here that, generally, a separation of at least $2\phi$ is required
to resolve the two microlensed images. The exact value of the minimal
separation probably lies in the range $\phi$ to $2\phi$, and depends on the
details of the procedure for faint source separation in the crowded inner
$1\arcsec$ of the GC.}, and the lensed source will appear as a {\em
microlensing} event. For a given angular resolution, there is a maximal
lens--source distance $d_\mu$ for microlensing,
\begin{equation} 
d_\mu = \frac{D_8^2}{M_{2.6}}\left(\frac{\phi}{0.018\arcsec}\right)^2\,
{\rm pc}\,.
\label{e:dmu}
\end{equation} 
More distant stars will have $\Oe > \phi$ and their two images will be
separately resolved.

\subsection{The microlensing rate}
\label{s:mlrate}

The unresolved images of a faint star at $d<d_\mu$ close to the line of
sight will appear as a `new source' at the position of $\SgrA$, to within
the observational resolution.  The total source amplification, $A$, is
related to $u$ by (\cite{Paczynski86})
\begin{equation}
u^2 = 2A/\sqrt{A^2-1}-2\,,
\end{equation}
and for small $u$, $A \sim 1/u$.  The maximal amplification along the
star's trajectory is reached when its projected position is closest to the
lens. A star of stellar type $s$ and an absolute $K$ magnitude $K_s$, which
is observed on a line of sight with an extinction coefficient $A_K$ and
distance modulus $\Delta$, will be detected above the flux threshold $K_0$
if it is amplified by at least $A_s=10^{-0.4(K_0-K_s-\Delta-A_K)}$. This
corresponds to a maximal impact parameter of $u_{0,s}$. As the minimal
required amplification approaches 1, $u_0$ diverges. This, and the fact
that in practice the threshold is not sharply defined, leads us to
introduce a cutoff at $u_0=1$, which implies a minimal amplification of
$A=1.34$.

The basic quantity that is required for predicting the lensing properties
is the differential lensing rate as function of stellar type, event
duration, amplification, and source distance from the black hole. We derive
an explicit expression for the differential lensing rate in the appendix.
Here it is instructive to discuss the relations between these properties by
considering the simpler case of the total lensing rate, regardless of the
event duration. The total integrated rate for microlensing amplification of
background stars above the detection threshold is
\begin{equation}
\Gamma(K_0) = 2\avu \int_{r_1}^{r_2} \Re \avv n_\star\,dr\,,
\label{e:rate}
\end{equation}
where $r_1$ is the inner radius of the central stellar cluster, $r_2$ the
maximal radius for producing unresolved lensed images, $\avv$ is the
transverse 2D stellar velocity averaged over the velocity distribution
function, and $n_\star$ is the total number density of stars. Here and
below, the notation $\langle\ldots\rangle$ designates the average of the
bracketed quantity over the stellar types with $K_s>K_0$, weighted by their
fraction in the stellar population, $f_s$, where it is assumed that $f_s$
does not depend on $r$. The properties of the stellar population enter the
integrated rate only through the mean impact parameter $\avu$, which for
$u_0
\ll 1$ is simply
\begin{eqnarray}
\avu & = & \sum_{\{s | K_s>K_0\}} f_s u_{0,s} \nonumber\\
& \simeq & \sum_{\{s | K_s>K_0\}} f_s A_s^{-1} = \langle F_K\rangle/F_0\,,
\label{e:avu}
\end{eqnarray}
where $F_K$ is the observed (dust extinguished) stellar $K$-band flux and
$F_0$ is the detection threshold flux corresponding to $K_0$.  

We characterize the microlensing time-scale for stars of type $s$ as the
average time they spend {\em above the detection threshold},
\begin{equation} 
\bar{\tau}(K_0) = \frac{\pi}{2}u_{0,s}\Rs \avv\,,
\label{e:t0}
\end{equation} 
where a $\pi/4$ factor comes from averaging over all impact parameters with
$u\le u_{0,s}$.

The lensing rate, amplification and event duration are inter-related. For
$u_0 \ll 1$, the {\em median} value of the distribution of maximal
amplifications is simply $A(u_0/2) \simeq 2A(u_0)$. Note, however, that the
{\em mean} maximal amplification, $\langle A_{\rm max}\rangle \simeq
\int_0^{u_0} (u_0 u)^{-1}\,du$, diverges logarithmically.  Generally, a
fraction $x$ of the events will have a maximal amplification of $A_0/x$ or
more (i.e. a peak magnitude above the threshold of $\DK = 2.5\log x$ or
less). The rate of such events is smaller, 
\begin{equation}
\Gamma(x) = x\Gamma(K_0)\,.
\label{e:qc}
\end{equation}
The time-scale of events amplified by a factor of $1/x$ above the threshold
is somewhat longer than the average time-scale (Eq.~\ref{e:t0}), since they
must cross closer to the line of sight,
\begin{equation}
\bar{\tau}(x) =
\frac{2}{\pi}\left(\sqrt{1-x^2}+\frac{\sin^{-1}x}{x}\right)\bar{\tau}(K_0)\,,
\label{e:t0c}
\end{equation}
which approaches $4\bar{\tau}(K_0)/\pi$ for large
amplifications. Conversely, for a given $v_2$ and $u_{0,s}$, even a small
increase in $\bar{\tau}$ is associated with a considerable increase in the
maximal amplification.

\subsection{Resolved lensed images}

When $\Oe > \phi$ at $d > d_\mu$, the two images can be resolved. As we
show in \S\ref{s:results}, $d_\mu$ is already quite small for present-day
angular resolutions, and will become smaller still as the resolution
improves. This implies that there may be a non-negligible contribution to
the lensing rate from regions beyond $d_\mu$. We therefore have to consider
also the case of resolved images.

For $u \ll 1$, the two images will appear at an offset of $\sim \Oe$ from
$\SgrA$, on opposite sides of it. The amplifications of the individual
images are related by $A = A_1+A_2$ and $A_2 = A_1+1$, which for the high
amplifications that are requires in the GC can be approximated as $A_1
\approx A_2 = A/2$. The formalism used for calculating the lensing of
unresolved images can therefore be applied in this case simply by raising
the effective detection threshold by a factor of two ($0.75\mag$). This
makes resolved images harder to detect, but on the other hand, if both
images are observed, the identification of the event as lensing is much more
certain.

\subsection{Lensing of observed bright stars}

The formalism for describing unresolved and resolved lensing of stars from
below the detection threshold can be also extended to cases where the
microlensed source is an observed bright, $K<K_0$ star. At the current
detection threshold, the observed central surface density is $\sim10$
arcsec$^{-2}$ (\cite{Genzel97}; \cite{Ghez98}). For such a small number of
stars, whose orbits can be tracked individually, a statistical treatment is
not very useful. However, our calculations indicate that a ten-fold
improvement in the detection threshold will yield an observed central
surface density of at least $\sim100$ arcsec$^{-2}$ (Eq.~\ref{e:Dp}). For
such a large sample, a statistical treatment is more meaningful. We
therefore present below, for completeness, results for microlensing
amplification of stars fainter than the detection threshold (`faint-star
lensing') as well as for stars brighter than the detection threshold
(`bright-star lensing'). For bright-star microlensing we assume that the
effective minimum magnification factor for detection is $A=1.34$ ($u_0=1$).

\section{Modeling the stellar population in the Galactic Center}
\label{s:gc}

Three basic quantities enter into the computation of the microlensing rate:
the stellar number density distribution, the stellar velocity field, and
the $K$-band luminosity function (KLF). 

The stellar population in the central $\sim 100$ pc appears to consist of a
mixture of old bulge stars, and `central cluster' stars which may have been
produced in various star-formation episodes during the lifetime of the
Galaxy (\cite{GHT94}; \cite{SM96}). Evidence for recent star formation in
the GC has been provided by Krabbe et al. (1995) who found a concentration
of luminous early-type stars within a few arcsec of $\SgrA$, implying a
recent ($\la 10^7$ yr) starburst in which $\sim 10^{3.5}$ stars were
formed.  Additional young stellar systems, the `Arches' and `Quintuplet'
clusters also exist close the GC and contain large numbers of massive stars
(\cite{SSF98}).  A further indication of on-going star-formation in the GC
is the fact that the KLF in the central cluster extends to more luminous
stars than the KLF of the Galactic bulge as observed via Baade's window
(\cite{LR87}; \cite{Tiede95}; \cite{Blum96}; \cite{Davidge97}).  Serabyn \&
Morris (1996) suggested that continuous star formation in the central
cluster is maintained by molecular clouds in the GC, and that the $\sim
r^{-2}$ radial light profile of the central cluster reflects the
distribution of the star-forming molecular clouds.

In our analysis we make the simplifying assumption that the discrete young
clusters in the GC can be modeled by a volume averaged and smoothly
distributed stellar population. In particular, we do not consider here
lensing events that might be associated with the innermost stellar cusp,
which has been identified by Eckart \& Genzel (1997) and Ghez et al.
(1998) in the immediate vicinity of $\SgrA$. An over-density of stars above
the smoothed distribution very near the black hole may contribute to very
short duration microlensing events. However, the total stellar mass and
luminosity function of the stars that are associated with this `$\SgrA$
cusp' are poorly constrained at the present time, so that computations of
the lensing rates require more extensive modeling and analysis. We will
present such computations elsewhere.

Because the lensing time-scales increase with distance, long duration events,
which are relevant for low sampling rates of the current proper motion studies,
tend to be associated with more distant stars.  We therefore consider both
stars in the central star-forming cluster, as well as more distant
old-population bulge stars in our analysis.  We now proceed to discuss the
stellar densities, velocities and KLFs of these two components.

\subsection{Stellar densities and velocities}

Genzel et al. (1996, 1997) derived density and velocity models for the
central cluster based on fits to the observed star counts, stellar radial
velocities and proper motions in the inner few pc. Their mass density
distribution is parameterized by a softened isothermal distribution
\begin{equation}
\rho_{\rm core}(r) = \frac{\rho_c}{1+3(r/r_c)^2}
\label{e:nstar}
\end{equation}
Where $r_c$ is the core radius and $\rho_c$ is the central density, with
best fit values of $r_c = 0.38\,$pc and $\rho_c = 4\,10^6\,\Mo{\rm
pc}^{-3}$. The 1D velocity dispersion of the late type stars in the core,
which are dynamically relaxed, is modeled by (\cite{Genzel96})
\begin{equation}
\sigma_{\rm core} = (55^2+103^2(r/r_{10})^{-2\alpha})^{1/2}
\,\,{\rm km\,s^{-1}}\,,
\end{equation}
where $\alpha = 0.6$ and $r_{10}$ is the projected distance corresponding to
$10\arcsec$.

In the inner GC, the mass density is strongly constrained by the dynamics,
whereas the $K$ luminosity density $\nuk$ ($L_{K\odot}\,{\rm pc}^{-3}$) is
not well defined due to the patchy nature of the extinction. The situation
is reversed on scales larger than 1\,kpc, where the observed rotation curve
and velocity dispersion are harder to interpret, but $\nuk$ can be
de-projected from the observed surface brightness. Kent (1992) proposed
that both the rotation curve and the surface brightness along the major
axis of the Galactic bulge in the mid-plane of the galaxy at $r\ga1\,$kpc
can be described as a superposition of bulge and disk components with a
mass-to-light ratio\footnote{Following Kent (1992), we define the
mass-to-light ratio as $\Upsilon = (M/\Mo)/(\LK/L_{K\odot})$, where the
solar monochromatic luminosity at $2.2\mu$m is $L_{K\odot} =
2.154\,10^{32}\,$erg\,s$^{-1}$\,$\mu$m$^{-1}$. Note that Genzel et
al. (1996) define this quantity as $\Upsilon^\prime = (M/\Mo)/(\lambda
L_\lambda/\Lo)$, with $\lambda = 2.2\mu$m. The two definitions are related
by $\Upsilon^\prime = 8.07\Upsilon$.} $\Upsilon = 1\MLKo$ and a luminosity
density
\begin{eqnarray} 
\nuk(r) & = & \nuk_{\rm bulge}+\nuk_{\rm disk}\\ \nonumber
        & = & 3.53K_0\left(\frac{r}{r_b}\right) +
            3\exp\left(-\frac{r}{r_d}\right)\,\,L_{K\odot}{\rm pc}^{-3}\,,
\end{eqnarray} 
where here $K_0$ is a modified Bessel function (not to be confused with the
detection threshold), $r_b = 667\,$pc and $r_d = 3001\,$pc. Kent also
suggested a $\nuk \propto r^{-1.85}$ extrapolation of $\nuk$ towards the
center based on observations of the intensity variation in the inner 10
pc. We replace this extrapolation with an updated and non-divergent model
by adopting the Genzel et al. (1996) mass density model (Eq.~\ref{e:nstar})
for the core and by assuming $\Upsilon = 1\MLKo$ for the bulge and disk, so
that $\rho_{\rm bulge} = \nuk_{\rm bulge}$ and $\rho_{\rm disk}= \nuk_{\rm
disk}$. We further assume that the bulge is axisymmetric, and model the
mass density over the entire distance range by
\begin{equation} 
\rho = \rho_{\rm core} + \rho_{\rm bulge}+\rho_{\rm disk}\,.
\label{e:rho}
\end{equation} 
Figure \ref{f:rho} shows our mass density model, and in particular the
emergence of the bulge component at $\sim 100$ pc. The central cluster
completely dominates the total mass density in the inner 10 pc, but then
falls to $\sim85$\% of the total at 50\,pc and to $\sim65$\% at $100\,$pc.


The velocity field on this scale includes both the bulk rotation and the
random motion. We approximate Kent's models for the galactic rotation and
the velocity dispersion in the inner 300\,pc along the major axis of the
bulge by log-linear fits. We assume that the rotation velocity is
perpendicular to the line of sight, and that its contribution to the
transverse velocity is
\begin{equation} 
v_{\rm rot} = 80+20\log_{10}(r/{\rm 1pc})\,\,{\rm km\,s^{-1}}\,,
\end{equation} 
and that the 1D velocity dispersion in the bulge is
\begin{equation} 
\sigma_{\rm bulge} = 60.9+18.9\log_{10}(r/{\rm 1pc})\,\,{\rm km\,s^{-1}}\,.
\end{equation} 
We model the 1D velocity dispersion over the entire distance range by
\begin{equation} 
\sigma = \max(\sigma_{\rm core},\sigma_{\rm bulge})\,.
\end{equation} 

Note that both the proper motion of $\SgrA$, which is $\la 20\,{\rm
km\,s}^{-1}$ (\cite{Backer96}), and the Sun's galactocentric rotation have
a negligible contribution to the relative proper motion of the source and
lens because of the high stellar velocities near the dynamic center and
because $d\ll D$.

\subsection{$K$ luminosity function}

In our analysis we consider a model for the KLF which is based
on the observed portions of the KLFs in the central cluster
and in the bulge, and on a theoretical KLF computed in a population
synthesis model for the central cluster.

\subsubsection{The power-law KLF}
\label{s:plklf}

The KLF of the bulge has been observed through Baade's window, which at
$l=1.0^\circ$, $b=-3.9^\circ$, is $\sim0.6\,$kpc from the GC at the
tangential point. The bulge KLF approximately follows a single power-law
$d\log N_\star/d\log\LK = \beta$, with $\beta = 1.695$ (\cite{Tiede95}),
from $M_K \sim -7.5\mag$ to $M_K \sim 2\mag$ \footnote{$M_K = -2.5\log \LK
+84.245$ mag for $\LK$ in erg\,$s^{-1}$\,$\mu{\rm m}^{-1}$.}, assuming a
distance modulus $\Delta=14.5$ and $A_K\sim0$. The faintest observed stars
are at the detection threshold, and the power-law KLF likely extends, or
steepens slightly, to lower luminosity stars near the main-sequence
turn-off point.

Recent {\em HST} observations of the optical luminosity function in Baade's
Window (\cite{Holtzman98}) probe it down to very low stellar masses ($\sim
0.3\, \Mo$), well below the main-sequence turn-off point. The $V$-band
luminosity function (VLF) presented by Holtzman et al. (1998) shows that a
sharp break occurs at the turn-off point $M_V \sim 4.5$--$5\mag$,
corresponding to $\sim 1\,\Mo$ stars.  We extrapolated the power-law KLF of
Tiede et al. (1995) down to $M_K = 3.5\mag$ (assuming $V$-$K \sim 1.5$ for
$1\,\Mo$ stars) and compared it to the Holtzman et al. (1998) VLF at $M_V =
5\mag$. We find that the predicted $K$ star-counts somewhat underestimate
the $V$ star-counts, but agree with them to within a factor of 2. We thus
conclude that the power-law KLF can be extended down to at least $M_K =
3.5\mag$. The observed VLF can be used to determine the behavior of the KLF
at even lower luminosities by using the $V$-to-$K$ conversions for low mass
stars presented by Henry \& McCarthy (1993). Doing this shows that the KLF
likely flattens at a break point close to $M_K=3.5\mag$, with $\beta \sim
1.5$ in the range $3.5\mag\la M_K \la 6.5\mag$, and turns over at
$M_K\sim7\mag$.

As we have discussed above, the stellar population in the central cluster
appears to be consistent with continuous star-formation throughout the
Galaxy lifetime (see e.g. Serabyn \& Morris 1996).  However, despite the
many differences between the populations in the bulge and the central
cluster, the observed KLF in the central $12.4\arcsec \times 11.9\arcsec$
of the GC follows a power-law similar to that of the bulge KLF
(\cite{Davidge97}), but extends to more luminous stars (\cite{Blum96}).
The exact upper luminosity cutoff is not well-defined due to statistical
fluctuations in the counts and the contribution of asymptotic giant branch
(AGB) stars, which do not follow the power-law.

The central cluster's KLF is not known beyond the current detection
threshold of $K_0 \sim 17\mag$, which is the region of interest for the
microlensing calculations. However, the similarity of the KLF power-law
indices in the central cluster and bulge KLFs suggests that the two KLFs
are similar, apart for having different upper luminosity cutoffs. The much
lower extinction in Baade's window, which is only $A_K=0.14\mag$ as
compared to $3.4\mag$ in the GC (\cite{RRP89}), makes it possible to extend
the KLF in the GC to lower luminosities. Indeed, Blum et al. (1996) find
that the bulge KLF, after correcting for the $A_K$ difference, can be
smoothly joined to the GC LF in the central $2^{\prime} \times
2^{\prime}$. Their resulting composite LF has a best fit power-law index
$\beta = 1.875$, and extends from $M_K\sim-10\mag$ down to $M_K =
2\mag$. Our theoretical model for the central cluster, which we discuss
below, indicates that this power-law character is further maintained down
to $M_K \sim 3.5\mag$.

We therefore adopt the same $\beta = 1.875$ power-law index for both the
central cluster and bulge KLFs.  For the central cluster KLF we set the
upper luminosity cutoff at $K_h=8\mag$, and for the bulge KLF at
$K_h=10.5\mag$. We set an effective low luminosity cutoff for both at
$K_l=21.5\mag$, which corresponds to stars with masses $\sim 1\Mo$.  As we
show in \S\ref{s:fit}, the lensing rates are insensitive to the exact
choice of the low luminosity cutoff.

We note that the observed bulge KLF is better fitted by a somewhat flatter
power-law than $\beta = 1.875$, which implies less faint-star lensing
candidates. On the other hand, the comparison with the optical luminosity
function suggests that the KLF is a conservative estimate of the total
number of stars. In addition, the flatter power-law fails to account for
the strong excess of horizontal branch (HB) stars above the power-law
(\cite{Tiede95}).  These are important potential microlensing sources, as
they lie in a magnitude range that is just below the threshold if the bulge
population is observed through the GC extinction of $A_K=3.4\mag$.  We
therefore consider these small discrepancies in slope and number to be
within the limitations of the power-law approximation and the observational
uncertainties.

\subsubsection{The theoretical KLF}
\label{s:popsynt}

As a check on the empirically based pure power-law KLF, we have also
computed a series of theoretical KLFs using our `population synthesis' code
(see e.g. \cite{Sternberg98}).  

In our models we used the Geneva stellar evolutionary tracks
(\cite{Schaerer93}), and concentrated on stellar models with twice-solar
metallicities, as is indicated by the enhanced abundances of the gas in the
GC (\cite{MS96}).  However, recent measurements of stellar spectra of cool
luminous stars in the GC point to solar metallicities
(\cite{Ramirez98}). We therefore considered also such stellar models, and
verified that our results do not depend strongly on the assumed
metallicity.  We computed the stellar $K$ luminosities using the empirical
bolometric corrections and $V$-$K$ colors for dwarfs, giants and
supergiants compiled by Schmidt-Kaler (1982), and Tokunaga (1998). The
Geneva tracks for intermediate mass stars ($\sim 2$--$7 \Mo$) do not extend
beyond the end of the early AGB phase. We extended these tracks to include
also the more luminous thermal-pulsing AGB phases following prescriptions
described by Charlot \& Bruzual (1991) and Bedijn (1988).  In our models we
assume explicitly that stellar remnants and stars less massive than
$0.8\Mo$ are negligible sources of $K$-band luminosity. This low-mass
luminosity cutoff roughly corresponds to the low luminosity cutoff of the
empirically based power-law KLF.


We constructed theoretical models for a range of cluster parameters
assuming continuous star-formation lasting for 10\,Gyr. We considered
models with pure power-law IMFs, and the Miller-Scalo IMF (\cite{MS79};
\cite{Scalo86}) for a range of lower- and upper-mass cutoffs.
We selected the model which yields a KLF which best matches the observed
KLF for the central cluster measured by Davidge et al. (1997).

We find that a model with a Miller-Scalo IMF ranging from $0.1$ to
$120\,\Mo$ provides the best overall fit to the data. In Figure~\ref{f:klf}
we compare our model KLF for the central cluster with the $\beta = 1.875$
power-law fit of Blum et al. (1996) to their composite KLF.  The overall
agreement between the theoretical KLF and the power-law KLF is remarkable,
for both stellar models with solar and twice-solar metallicities. The model
successfully reproduces both the power-law character and slope of the
observed KLF of the central cluster. Furthermore the model shows that the
power-law KLF likely extends down to at least $K=21.5\mag$. In this model,
the most probable lensed sources are $K\sim 21\mag$ stars with $M_\star
\sim 1\,\Mo$, $R_\star\sim 10^{11}$\,cm just off the main sequence, which
require a magnification of order $A\sim 50$ to be detected above the
threshold.

Our population synthesis model predicts a mass-to-light ratio $\Upsilon =
0.24\MLKo$, in excellent agreement with $\Upsilon \sim 0.25\,\MLKo$
measured in the central cluster by Genzel et al. (1996).

\subsubsection{Normalizing the KLF}
\label{s:fit}

As shown by eq.~\ref{e:rate}, the lensing rate depends on, $n_\star$, the
total number density of stars which are effective sources of K-band
luminosity.  We refer to such stars as `K-emitting' stars, and define
\begin{equation} 
n_\star = \frac{f_\star}{\avm}\rho\,,
\label{e:norm-rho} 
\end{equation} 
where $\rho$ is the total dynamical mass density, $\avm$ is the mean
stellar mass of the K-emitters, and $f_\star$ is the fraction of the total
dynamical mass contained in $K$-emitting stars, i.e. excluding
low-mass stars and remnants (i.e. objects which lie below the effective
low-luminosity cutoff of the KLF) and gas clouds.

The observed star counts, $dN^{\rm obs}_\star/d\LK$, within an area S, can
be used to obtain a best fit value of $f_\star/\avm$ given the constraint
\begin{equation} 
\frac{dN^{\rm obs}_\star}{d\LK} = \frac{df}{d\LK} 
\frac{f_\star}{\avm}\int\rho\,dS\,dr\,,
\label{e:norm}
\end{equation}
where the KLF $df/d\LK$ is normalized to 1. The mass-to-light ratio over
the integration volume used in Eq.~\ref{e:norm} can then be deduced from
the fit by
\begin{equation} 
\Upsilon = \frac{M}{\LK} = \frac{\avm}{f_\star\avL}\,.
\label{e:mtol}
\end{equation} 
where $\avL$ is the average K-band luminosity of the K-emitting stars.

The power-law KLF we adopt in our computations
($df/d\LK\propto\LK^{-\beta}$) is characterized by $1<\beta<2$, $\La \ll
\Lb$ and $\La\ll \Lt \ll \Lb$, where $\La$ and $\Lb$ are the effective
lower and upper luminosity cutoffs to the KLF, and $\Lt$ is the $K$
luminosity corresponding to the detection threshold. By approximating $u
\sim 1/A$ (where $A$ is the required amplification) the mean stellar
luminosity and mean impact parameter for such KLFs are given by the
simple approximate relations
\begin{eqnarray} 
\avL & \approx &
\frac{(\beta-1)}{(2-\beta)}\frac{\Lb^{2-\beta}}{\La^{1-\beta}}\,,\nonumber\\
\avu & \approx &
\frac{(\beta-1)}{(2-\beta)}\left(\frac{\La}{\Lt}\right)^{\beta-1}\,.
\end{eqnarray}
It also follows that 
$n_\star$, $\Upsilon$ and the lensing rate $\Gamma$ scale as
\begin{eqnarray}
n_\star & \sim & f_\star/\avm \sim \frac{1}{\beta-1}\La^{1-\beta}\,,\nonumber\\
\Upsilon & \sim & (2-\beta)\Lb^{\beta-2}\,,\nonumber\\
\Gamma & \sim & \avu n_\star \sim
\frac{1}{2-\beta}\Lt^{1-\beta}
\,.
\label{e:plrho}
\end{eqnarray}
Eq.~\ref{e:plrho} shows that the total lensing rate is {\it insensitive to
the low-luminosity cutoff} of the KLF. This simply reflects the fact that
an increase in the number of $K$-emitting stars as $\La$ decreases is
offset by a correspondingly smaller mean lensing impact parameter for the
stellar system, and vice-versa. This important property allows us to
robustly compute the lensing rate even if the effective lower luminosity
cutoff of the power-law KLF is not well determined\footnote{We note that if
$\beta>2$ then the rate does depend on $\La$ with $\Gamma \sim \La^{2-\beta}$.
However, as we have discussed, the observations of Holtzman et al. (1998)
strongly suggest that the KLF flattens, rather than steepens, below our
assumed value for $\La$.}.

We note that since the differential contribution of stars with luminosity
$L_{K,s}$ to the mean impact parameter $\avu$ scales like
$(L_{K,s}/L_0)L_{K,s}^{-\beta} \sim L_{K,s}^{1-\beta}$, the integrated
contribution of stars from a $1\mag$-wide bin is, for $\beta=1.875$, a very
weakly decreasing function of the bin's $K$ magnitude.  We therefore expect
that the lensed stars will exhibit a wide range of intrinsic $K$
magnitudes, with a weak trend towards lensing by stars close to the
detection threshold.

It is uncertain at which radius the stellar population makes the transition
from a population that is characteristic of a star-forming cluster to a
more bulge-like population. However, as we argued above, the observed
properties of the KLFs in the GC and the bulge suggest that they are very
similar for $K>10.5\mag$. Since the normalization of the KLF does not
depend strongly on the upper luminosity cutoff (Eq.~\ref{e:plrho}), and
since the very high luminosity tail of the KLF is not relevant for the
lensing calculations (the $8\mag<K<10.5\mag$ stars have a negligible
contribution to the lensing rate of observed stars), we infer the value of
$f_\star/\avm$ using the star-counts observed in the core, and we adopt
this normalization for the entire inner 300 pc.

In carrying out this procedure\footnote{The volume integration in
Eq.~\ref{e:norm} was carried out by approximating the rectangular field
with a circular field of projected radius $p$ having the same area. The
integration was done over a cylinder of radius $p$, centered on the black
hole and extending along the line of sight $300\,$pc in each direction, a
distance large enough to ensure that the surface density reaches its
asymptotic value.} we used the observed number counts in the $K=14\mag$ bin
(stars per mag per arcsec$^{-2}$), averaged over the $12.4\arcsec\times
11.9\arcsec$ field observed by Davidge et al. (1997). The star counts in
this luminosity range are likely complete, and this range is also well
separated from the regions that are affected by AGB and HB stars, which
cause deviations from the power-law behavior (\cite{Tiede95}). Using
Eq.~\ref{e:norm}, we then find that $f_\star/\avm = 0.2\,\Mo^{-1}$. This
value for $f_\star/\avm$ can be reproduced, for example, by a choice of
$f_\star = 0.2$ and $\avm = 1\Mo$, which is comparable with the values
$f_\star=0.22$ and $\avm = 0.84\Mo$ of our population synthesis model for
the central cluster.  For the central cluster power-law KLF,
$\avL=22\,L_{K\odot}$, so that Eq.~\ref{e:mtol} yields $\Upsilon \sim
0.22\,\MLKo$, again in excellent agreement with $\Upsilon \sim 0.25\,\MLKo$
inferred by Genzel et al. (1996) for the central cluster.

\section{Results}
\label{s:results}

We have carried out detailed computations of the lensing event rates using
the formalism described in \S\ref{s:rate} and in the appendix, and using
the stellar number and velocity distributions and the power-law $K$-band
luminosity function, as discussed in \S\ref{s:gc}.  The integrations were
carried out from a minimum distance $r_1 = 0.005\,$pc, equal to the mean
central stellar separation, to a distance $r_2 = 300\,$pc, where the
integrated lensing rates approach their asymptotic values.

The normalized central cluster KLF allows us to estimate the stellar
surface density as a function of the detection threshold, and from it
derive the mean angular separation of the stars, $\Dp$, the required
angular resolution of the observations $\phi$, and the maximal distance for
observing (unresolved) microlensed stars $d_\mu$ (as given by
Eqs.~\ref{e:phi} and \ref{e:dmu}).  We find that for the central cluster
KLF
\begin{eqnarray} 
\log \Dp & = & 2.31-0.175 K_0\,,\nonumber\\
\log S_0 & = & -4.62+0.35 K_0\,,
\label{e:Dp}
\end{eqnarray} 
for $\Dp$ in arcsec and $S_0$ in arcsec$^{-2}$. Thus, for a detection
threshold of $K_0 = 17\mag$ we expect a central surface density of 21 stars
per arcsec$^{-2}$ for complete counts\footnote{Genzel et al (1997) and Ghez
et al (1998) reported $S_0\sim20$ and 15 stars arcsec$^{-2}$,
respectively. However, the star counts are incomplete close to the detection
thresholds, and the measured $S_0$ actually indicate a significant density
enhancement (the `$\SgrA$ cusp') relative to the immediate surroundings.}.

We present our results in a way which makes it possible to flexibly
estimate the actual detection probabilities for a broad range of observing
strategies.  We consider the generic monitoring campaign to consist of a
series of $n$ very short observing runs which are carried out during a
total time period $T$ (typically $T \sim$ several years), with a mean
interval $\Delta T$ between each observing run (typically $\DT \sim$ a
month to a year).  We now wish to distinguish between `long' and `short'
events, where we define the event duration as the time the source spends
above the detection threshold. Long events are those with durations $\tau >
\DT$, and would appear as time varying sources that brighten and fade
during the course of several observing runs.  The microlensing origin of
long events could be verified, in principle, from the shape of the light
curve and its achromatic behavior. Short events are those with durations
$\tau < \DT$, and would be observed as a single `flash' provided they occur
within a time $\tau$ of any one of the $n$ observing runs. In the limit of
small event rates the detection probability may be written as
\begin{equation} 
P = P_{\rm short}+P_{\rm long} = n\bar{\tau}_{\rm short} \Gamma_{\rm short}
+ T\Gamma_{\rm long\,,
\label{e:P}}
\end{equation} 
where $\Gamma_{\rm short}$ and $\Gamma_{\rm long}$ are the rates of short
and long events, and $\bar{\tau}_{\rm short}$ is the rate averaged lensing
duration for short events. In the (ideal) limit of continuous monitoring,
$\Gamma_{\rm long}$ approaches the total event rate and $P = T\Gamma_{\rm
total}$.

The results of our computations are displayed in Figs.~\ref{f:cumr} and
\ref{f:dK}. In Fig.~\ref{f:cumr} we plot the cumulative rates,
$\Gamma_{\rm long} (\tau>\DT)$, for all lensing events with durations
$\tau$ longer than the timescale $\DT$, as functions of $\DT$.  We present
results for events which produce unresolved and resolved images for stars
which are either intrinsically below or above the detection thresholds. We
present computations for detection thresholds ranging from 16 to 19
magnitudes.  The total lensing rates, can be read off the plot from the
asymptotic values of the curves as $\Delta T \rightarrow 0$.  The curves
are flat for timescales less than $\sim$ 1 week, which is shorter than most
of the event durations.  As $\DT$ increases $\Gamma_{\rm long}$ decreases
as a smaller fraction of the lensing events satisfy $\tau > \DT$.  As an
example, Fig.~\ref{f:cumr} shows that for $K_0=17\mag$, the total lensing
rate is equal to $3\times 10^{-3}$ yr$^{-1}$, and that for events with
durations greater than 1 yr the rate is equal to $1\times 10^{-3}$
yr$^{-1}$.

In Fig.~\ref{f:cumr} we also plot the rate-averaged lensing timescale,
$\bar{\tau}_{\rm short}$, for events with $\tau <\DT$, which is needed to
estimate the detection probability of short events. The values of
$\bar{\tau}_{\rm short}$ are almost independent of $K_0$, since the shape
of the cumulative rate function is insensitive to $K_0$. We note also that
for small timescales $\bar{\tau}_{\rm short} \approx \DT/2$, as would be
expected for a rate which is independent of the timescale. The rate of
short events is simply $\Gamma_{\rm short}(\DT) = \Gamma_{\rm
long}(0)-\Gamma_{\rm long}(\DT)$.  Thus, for $K_0=17\mag$ the rate of
events lasting less than 1 yr is $2 \times 10^{-3}$ yr$^{-1}$, and the
average duration of such events is about 3 months.

In Fig.~\ref{f:dK} we plot $\DK_{\rm long}$, the median value of the maximal
amplifications for long events ($\tau>\DT$), as a function of $\DT$. Long
events which amplify sub-threshold stars are associated with stars at
greater distances from the black hole, stars at the low velocity tail of
the velocity distributions, or stars that cross closer to the line of sight
to $\SgrA$. Because of the latter effect, $\Delta K_{\rm long}$ is greater
than $0.75\mag$, which is the median value for all the events. This effect
is less marked for resolved lensing, which is characterized by longer
timescales, and is very weak for lensing of stars which lie above the
detection threshold, where $\Delta K_{\rm long}$ approaches the limit $\sim
0.75\mag$.

Figure~\ref{f:dqdr} shows the contributions to the lensing rate from
different regions in the GC, for a specific example where $K_0=16.5\mag$
and $\DT = 1\,$yr. Several important features are apparent in the results
displayed in Fig.~\ref{f:dqdr} and \ref{f:cumr}. First, it is evident that
the cumulative rate $\Gamma_{\rm total}(<r)$ approaches its asymptotic
value at $r\la 10\,$pc, and that short lensing events due to stars near the
black hole dominate the total rate. For example, Fig.~\ref{f:cumr} shows
that the median timescale of unresolved amplifications of sub-threshold
stars for is $\sim 2$ months, and Fig.~\ref{f:dqdr} shows that the median
distance from the black hole is $\sim 0.3\,$pc.  The run of the
differential rate for long events, $d\Gamma_{\rm long}/dr$, with distance
illustrates that the inner regions hardly contribute any long events.
Second, unresolved lensing of sub-threshold stars have the shortest
timescales. This is because such events are due mainly to stars close to
the black hole where the velocities are high and the lensing cross sections
are small.  Unresolved lensing of stars which lie above the threshold, for
which the cross sections are larger, have somewhat longer timescales, and
resolved events, which are due to stars at larger distances from the black
hole, are longer still.  Third, the rates for resolved lensing are about an
order of magnitude smaller than those for unresolved lensing events.

We now apply our results to estimate the probability that the variable
$K$-band source (or sources) reported by Genzel et al. (1997) (source S12)
and Ghez et al. (1998) (source S3) was a microlensing event.  The
monitoring of proper motions in the GC has been going on for about $T =
6\,$yr, at a sampling interval of $\DT = 1\,$yr, with a detection threshold
of $K_0 = 16.5\mag$ and a FWHM resolution of $\la0.15\arcsec$
(\cite{Genzel97}). For this detection threshold, $\Delta p = 0.25\arcsec$
(which corresponds to a required spatial resolution $\phi = 0.15\arcsec$),
and $d_\mu = 65\,$pc. We can now use Figs.~\ref{f:cumr} and \ref{f:dK} and
Eq.~\ref{e:P} to estimate the detection probability and typical
amplification of a lensing event in this experiment. The rate of unresolved
and long events which amplify sub-threshold stars is $7.5\times
10^{-4}\,$yr$^{-1}$, with a resulting detection probability $P_{\rm long}
\sim 0.5\%$. The median amplifications of such events is $\Delta K_{\rm
long}\sim 1.5\mag$ above the detection threshold. The probability for
detecting a short amplification of sub-threshold stars is $P_{\rm
short}\sim0.2\%$. The probabilities for detecting unresolved events
involving above threshold stars is of the same order of magnitude. The
probabilities for detecting long resolved events are an order of magnitude
smaller, and the probabilities for detecting short resolved events are
negligible in this experiment. Thus, we find that the behavior of the
variable source (or sources) at $\SgrA$, i.e. a brightening of a previously
undetected source to $1.5\mag$--$2\mag$ above the threshold with an event
duration of $\sim 1$ yr, is the typical behavior that would be expected for
a microlensing event. However, we also find that the a-posteriori
probability for detecting such an event is only $\sim 0.5\%$.

The probability for detecting a lensing event can be increased considerably
by carrying out more sensitive observations at higher sampling rates. For
example, 10 years of monthly observations with a detection threshold of
$K_0 = 19\mag$ will require a resolution of $\phi = 0.06\arcsec$, which is
already available (\cite{Ghez98}), and will increase the total detection
probability of long lensing events to $P_{\rm long}\sim 20\%$. These
estimates are uncertain by a factor of few due to uncertainties in the
stellar density distribution, the $K$ luminosity function and the
extinction.

\section{Discussion and summary}
\label{s:discuss}

In the past several years, very high resolution, very deep IR
observations of the GC have regularly monitored stellar motions within few
arcsec of the radio source $\SgrA$. The primary objective of these efforts
is to dynamically weigh the central dark mass and set lower bounds on its
compactness. As a by product, these observations can detect and record the
light curves of faint time-varying sources in the inner GC over timescales
of years. These proper motion studies have recently revealed the possible
presence of one or two variable $K$-band sources very close to, or
coincident with, the position of $\SgrA$ (\cite{Genzel97}; \cite{Ghez98}).

The first issue to resolve, when considering results from such
technologically challenging observations, is whether these sources are
real, or merely artifacts of the complex procedures for obtaining deep
diffraction limited images in the crowded GC. Assuming that such a source
is indeed real, and that it is not simply a variable star that happens to
lie very near to the line of sight to $\SgrA$, there are two interesting
possibilities to consider, both directly linked to the existence of a
super-massive compact object in the GC. A time variable IR source,
coincident with $\SgrA$, may be the IR counterpart of the radio source,
with the IR flare resulting from fluctuations in the accretion process,
which has up to now eluded detection in any band other than the radio (see
e.g. \cite{Backer96}). Another possibility is that the new source is a
faint star in the dense stellar cluster in the GC, which is microlensed by
the black hole.  Such amplification events would appear as time varying
sources very close to the position of $\SgrA$.

We note that a detection of microlensing could provide an independent probe
of the compactness of the central dark mass.  The innermost observed stars
in the GC set an upper limit on the size of the dark mass, $r_{\rm dm}\sim
7\,10^4\,R_\bullet$, where $R_\bullet$ is the Schwarzschild radius
(\cite{Ghez98}).  Microlensing has the potential, given a well sampled
light curve, to improve on the dynamical limit, since for typical values of
the lens--source distance, $r_{\rm dm}\ll\Re < 3\,10^4\,R_\bullet$ (see
Eq.~\ref{e:Rsch}). If the central mass is not a black hole, but rather an
extended object, then the microlensing light curve will deviate from that
of a point-mass lens when $r_{\rm dm} \ga \Re$. Because the innermost
observed stars already constrain $\theta_{\rm dm} \la 0.1\arcsec$, only
lensed stars in the inner few pc have a small enough $\Oe$ to probe
possible structure in the dark mass distribution
(Eq.~\ref{e:Oe}). Therefore, the question whether there is a significant
chance to detect such events is related to the yet unresolved question of
whether there is a strongly peaked stellar cusp in the innermost GC. We
will discuss these issues elsewhere.

In this paper we investigated the possibility of microlensing amplification
of faint stars in the dense stellar cluster in the GC by the super-massive
black hole, which is thought to coincide with the radio source $\SgrA$. We
calculated in detail the rates, durations and amplifications of such
events, and considered separately the cases of unresolved and resolved
images and of intrinsically faint and bright sources. We presented our
results in a general way that can be used to estimate the detection
probabilities of microlensing for a wide range of observing strategies.

The background stellar surface density increases with the detection
threshold $K_0$. This determines the observational angular resolution 
required to detect a microlensing event, and 
fixes the maximal distance behind the black hole for which 
the two lensed images of a star will appear unresolved. This maximal
distance occurs before the integrated lensing rate reaches its asymptotic
value, even for present-day detection thresholds.  We therefore considered
two manifestations of the lensing: unresolved microlensing of stars near
the black hole, and resolved lensing of stars farther away. We find that
short lensing events, due to stars close to the black hole, dominate the
total lensing rate. This reflects the fact that the high stellar density
and velocities near the black hole more than compensate for the smaller
lensing cross-sections there. For this reason, and because unresolved
images are on average twice as bright as the resolved images, unresolved
microlensing dominates the lensing rate. We have also considered the
lensing amplification of bright observed stars. The contribution of this
type of microlensing to the total rate becomes progressively more important
as the detection threshold decreases, and at low sampling rates, which are
primarily sensitive to longer events. We find that low sampling rates
significantly bias the detection towards high amplitude events.

Our predicted lensing rates are small, but not so small as to be
negligible. In particular, longer, deeper proper motion monitoring done at
higher rates, e.g. 10 years of monthly monitoring with $K_0 = 19\mag$, may
have a significant chance of detecting such an event.

Finally, could either of the variable sources reported by Genzel et
al. (1997) (source S12) and by Ghez et al. (1998) (source S3) be the
amplified microlensed image of a faint star? The lack of evidence for
related variability in the radio and X-rays, as would be expected in some
accretion scenarios if the $K$-flare were due to fluctuations in the
accretion process, argues against the possibility that the new source is
the IR counterpart of $\SgrA$.

Another possibility is that the new sources are variable stars, which are
below the detection limit in their low state.  Long Period Variable stars
(LPVs), which are probably luminous Mira variables, are observed in the GC
(\cite{HR89}; \cite{Tamura96}).  Typical amplitude variations of $\DK\sim
0.15\mag$--$0.5\mag$ are observed over $1$--$2\,$ yr, although in some
cases the variations are as large as $\DK \ga 1\mag$. Haller \& Rieke
(1989) find 12 LPVs in a 4.5 arcmin$^2$ survey of the GC (not including the
central $1^{\prime}\times1^{\prime}$) down to $K = 12\mag$ ($M_K =
-5.9\mag$). Of these, only one exhibited high amplitude variations ($1\mag$
in 4 months). At $K\sim 15\mag$, the new variable $K$ source is much
fainter than the LPVs observed in this survey. If it is an intrinsically
bright LPV, it must be lie on a highly extinguished line of sight. Using
the observed surface density of $M_K<-5.9\mag$, $\DK\sim 1\mag$ LPVs, it is
possible to make a rough estimate of the probability for finding such a
star within $0.15\arcsec$ of $\SgrA$. Even after taking into account a
factor $\sim40$ difference in the surface mass density between the
dynamical center and the survey area at $\sim2.5^\prime$ from the center,
the probability is only $\sim 0.005\%$. This of course does not rule out
the possibility of an intrinsically faint but highly variable star. In any
case, if the new source is a variable star, future observations should
detect continued variability from this source.

We can also argue statistically against the possibility that these flares
are due to microlensing of a star by another star in the GC, which happens
to be close to the line-of-sight to $\SgrA$. Because
$\avm/M_\bullet\sim10^{-6}$, the typical lensing timescale by a star will
be of the order of $1$ hr, $10^3$ times shorter than that of lensing by the
black hole (Eqs.~\ref{e:Re} and \ref{e:rate}). Furthermore, it is
straightforward to show that the total rate of lensing of a star by a star
within an angle $0.1\arcsec$ of the line-of-sight to $\SgrA$ is $\sim
5\,10^{-5}$\, yr$^{-1}$ for $K_0 = 17\mag$, which is 100 time smaller than
the rate of lensing by the black hole.


Our analysis has shown that the behavior of the variable $K$-band source
(or sources) at $\SgrA$, in particular a brightening of a previously
undetected source to $\sim1.5$--$2\mag$ above the threshold, on a timescale
of $\sim 1$ yr, is the typical behavior that would be expected for a
microlensing event.  However, we also estimate that the probability that
such an event could have been observed during the course of the proper
motion studies that have been carried out thus far is only $\sim
0.5\%$. While this probability is small, it is not so small as to rule out
this possibility entirely. The probability of detecting a microlensing
event at $\SgrA$ will increase considerably in the future as the
observational sensitivities and the monitoring sampling rates improve.

\acknowledgments

We thank D. Backer, R. Blandford, A. Eckart, R. Genzel, A. Ghez, I. King,
E. Maoz, B. Paczy\'{n}ski and E. Serabyn for helpful discussions and
comments. This work was supported in part by a grant from the
German-Israeli Foundation (I-551-186.07/97). A. S. thanks the Radio
Astronomy Laboratory at U.C.  Berkeley, and the Center for Star Formation
Studies consortium for their hospitality and support.

\appendix

\section{The microlensing rate and amplification of long 
duration events}
\label{s:qtlong}

Figure~\ref{f:cumr} shows that the total lensing rate is dominated by short
events (shorter than a few month). Such events are shorter than the current
mean time between observations, and can therefore be observed only in the
rare cases where they occur simultaneously with an observing run
(\S\ref{s:rate}). Of more relevance are long events, which span two or
more observing runs. Longer events occur when the lensed star is either
slow, or when it is far away from the black hole, so that its trajectory
through the Einstein ring is long (Eq.~\ref{e:Re}), or when it passes close
to the optical axis (Eq.~\ref{e:t0c}). In this appendix we calculate the
rate of events that are longer than some specified time, and estimate their
median maximal amplification.

Microlensing events with $\tau>\DT$ are those with transverse velocity $v_2$
and impact parameter $u\Rs$ such that
\begin{equation} 
\tau = \frac{2\Rs\sqrt{u_{0,s}^2-u^2}}{v_2} > \DT\,,
\end{equation} 
where $s$ is the stellar type and $u_{0,s}$ is the maximal impact parameter
for amplification above the threshold. The local rate per stellar type is
\begin{equation} 
\frac{d^2\Gamma_2}{drds} = 
2\Rs n_\star \int_0^{u_{0,s}}du 
\int_0^{v_{\rm max}} v_2 {\rm DF_2}({\bf v}_2)\,d^2v_2\,,
\label{e:tmin}
\end{equation} 
where $v_{\rm max}= 2\sqrt{u_{0,s}^2-u^2}\Rs/\DT$ and $\rm DF_2$ is the 2D
distribution function of velocities. The 2D velocity ${\bf v}_2$ is
composed of an ordered component ${\bf v}_{\rm rot}$, which we assume to be
perpendicular to the line of sight, and a random isotropic component, which
we assume to be Gaussian with a 1D dispersion $\sigma$. The 2D distribution
function of the projected velocity in polar coordinates is
\begin{equation} 
{\rm DF_2}({\bf v}_2)dv_2d\theta = 
\frac{\tvt}{2\pi}\exp(-(\tvt^2+\tvrot^2-2\tvt\tvrot\cos\theta)/2)d\tvt\,d\theta\,,
\end{equation} 
where $\tvt=v_2/\sigma$, $\tvrot =v_{\rm rot}/\sigma$ and $\theta$ is the
angle between ${\bf v}_2$ and ${\bf v}_{\rm rot}$.  Upon substitution into
Eq.~\ref{e:tmin} and averaging over all stellar types, one obtains
\begin{equation} 
\frac{d\Gamma_2}{dr} = 2\Rs n_\star \sigma \langle u_0 G(\tvmax,\tvrot)\rangle\,,
\end{equation} 
where $\tvmax = 2u_{0,s}\Rs/\DT\sigma$, and
\begin{equation}
G(\tvmax,\tvrot) = \exp(-\tvrot^2/2)
\int_0^{\tvmax} x^2 \sqrt{1-(x/\tvmax)^2}\exp(-x^2/2)I_0(x\tvrot)\,dx\,,
\end{equation} 
where $I_n$ is a modified Bessel function of order $n$. In the simple case
where $v_{\rm rot} = 0$,
\begin{equation} 
G(\tvmax) = \frac{\pi}{2}\tvmax \exp(-\tvmax^2/4) I_1(\tvmax^2/4)\,.
\end{equation} 
In the limit of $\tvmax \rightarrow \infty$ ($\DT = 0$),
\begin{equation} 
G(\tvrot) = \sqrt{\frac{\pi}{8}} \exp(-\tvrot^2/4)
\left[(2+\tvrot^2)I_0(\tvrot^2/4)+\tvrot^2 I_1(\tvrot^2/4)\right]\,,
\end{equation} 
which is independent of $u_{0,s}$, so that 
$\langle u_0 G(\tvmax,\tvrot)\rangle\rightarrow\avu G(\tvrot)$.
It is straightforward to verify that in this limit, the mean transverse
velocity approaches its correct asymptotic values, since
\begin{eqnarray}
\lim_{\tvrot\rightarrow 0} \sigma G(\tvrot) & = & \sqrt{\pi/2} 
\sigma\,,\nonumber\\
\lim_{\tvrot\rightarrow \infty} \sigma G(\tvrot) & = & v_{\rm rot}\,.
\end{eqnarray}
A fraction of the long duration events are due to stars on trajectories
with smaller than average impact parameters. Therefore, the median impact
parameter of long events is smaller than $\avu/2$ and the median
amplification above the threshold is greater than 2. The local average
impact parameter is
\begin{equation} 
\bar{u}(r) = \langle\int_0^{u_0} du u \int_0^{v_{\rm max}} 
DF_2({\bf v}_2)\,d^2v_2 \rangle = \langle u_0 W(\tvmax,\tvrot)\rangle\,,
\end{equation} 
where the weight function $W$ can be written as
\begin{equation} 
W(\tvmax,\tvrot) = \exp(-\tvrot^2/2)
\frac{1}{2}\int_0^{\tvmax} x (1-(x/\tvmax)^2)\exp(-x^2/2)I_0(x\tvrot)\,dx\,.
\end{equation} 
It is straightforward to verify that 
\begin{equation} 
\lim_{\tvmax \rightarrow\infty} W(\tvmax,\tvrot) =\frac{1}{2}\,.
\end{equation} 
The total rate averaged impact parameter is
\begin{equation} 
\bar{u} = \left.\int_{r_1}^{r_2} \bar{u}(r) \frac{d\Gamma_2}{dr}\,dr\right/
\Gamma_2\,.
\end{equation} 
and the median amplification above the threshold can be estimated by $A =
\avu/\bar{u}$.

\clearpage


%
%
\clearpage
\section*{Figure captions}

\figcaption[rho.eps]{The GC mass density model.
\label{f:rho}}

\figcaption[klf.eps]{The KLF of the theoretical continuous star forming
model with twice solar metallicity (solid line) and solar metallicity
(dashed line), compared to the Blum et al. (1996) $\beta = 1.875 $
power-law model of the composite observed KLF (points).
\label{f:klf}}

\figcaption[dqdr.eps]{The differential and integrated microlensing rate 
as function of the distance behind the black hole for the central cluster
model, assuming $K_0 = 16.5\mag$ and both continuous monitoring and $\DT =
1\,$yr. Both the total rates and the rates for long duration events with
$\tau>\DT$ are shown. The discontinuity at $56\,$pc reflects the transition
from unresolved images to the fainter resolved images.
\label{f:dqdr}}

\figcaption[cumr.eps]{The cumulative lensing rate, 
$\Gamma_{\rm long} (\tau>\Delta T)$, and the rate averaged lensing
timescale $\bar{\tau}_{\rm short} (\tau<\Delta T)$ of the central cluster
KLF as function of the sampling interval $\Delta T$, for the detection
thresholds $K_0 = 16\mag$, $17\mag$, $18\mag$ and $19\mag$ (from top to
bottom). Bold lines show results for faint-star lensing and thin dotted
lines for bright-star lensing. The thin dashed line in the bottom
panels is $\bar{\tau}_{\rm short} = \Delta T/2$. (In the case of resolved
lensing, values of $\bar{\tau}_{\rm short} < 0.3\,$yr are not shown due to
numerical instabilities in the calculations.)
\label{f:cumr}}

\figcaption[dk.eps]{The mean amplification above the detection threshold,
$\Delta K$, as function of the sampling rate $\Delta T$. Bold lines show
results for faint-star lensing and thin dotted lines for bright-star
lensing.
\label{f:dK}}

%
%
\clearpage
\begin{figure*}
\centerline{\epsfxsize=6.0in\epsffile{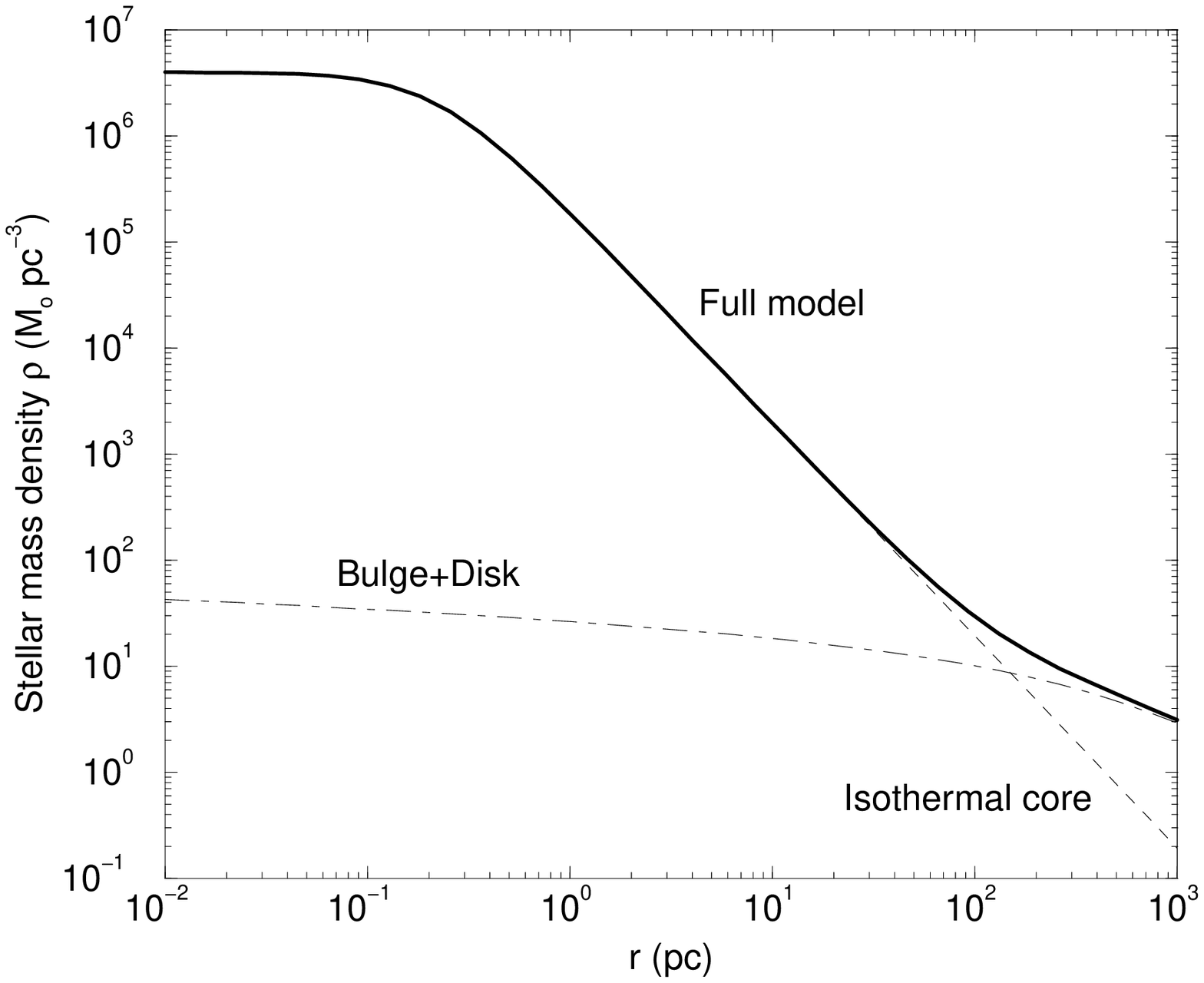}}
\centerline{Figure 1.}
\end{figure*}

\clearpage
\begin{figure}
\centerline{\epsfxsize=6.0in\epsffile{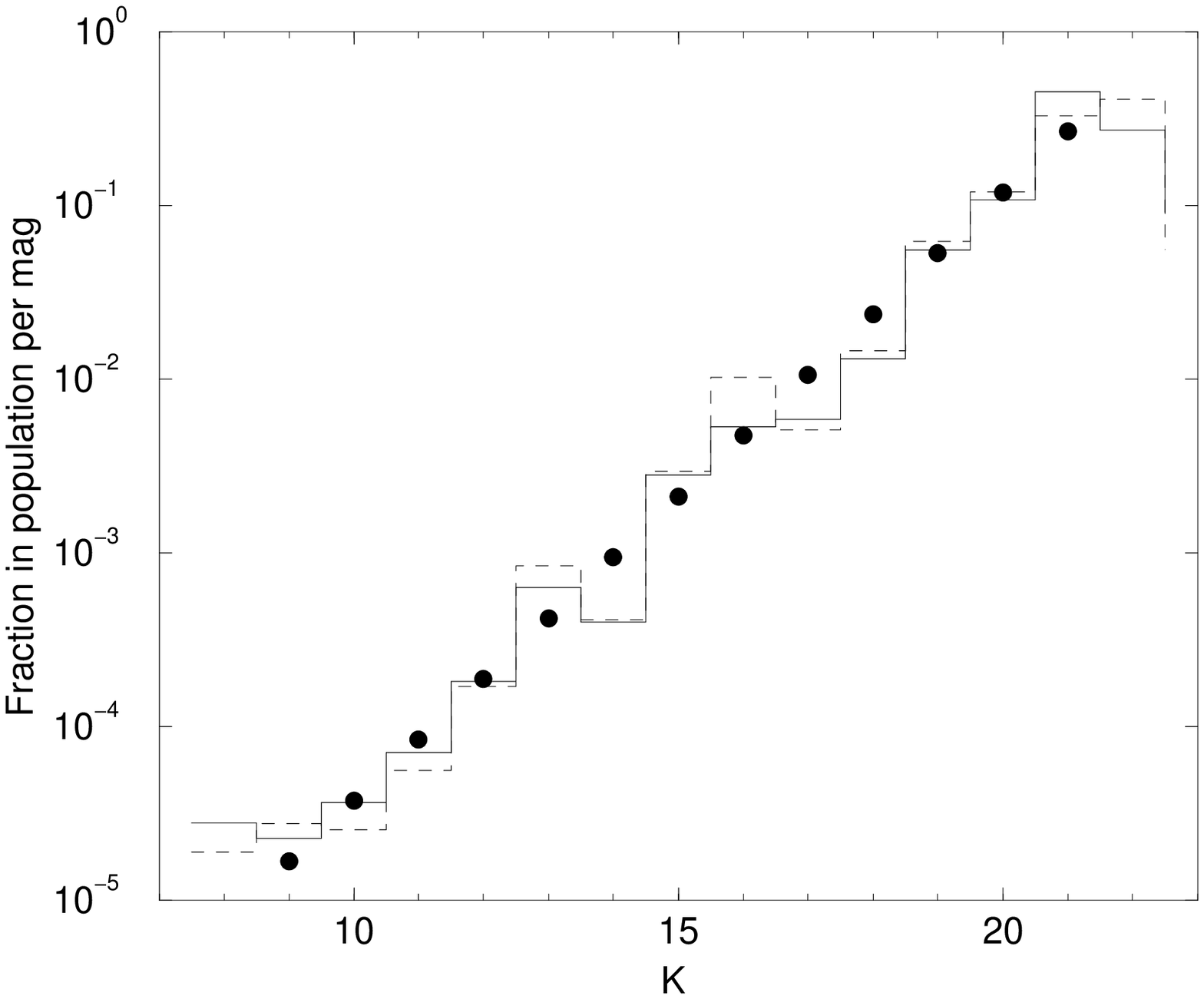}}
\centerline{Figure 2.}
\end{figure}

\clearpage
\begin{figure}
\centerline{\epsfxsize=6.0in\epsffile{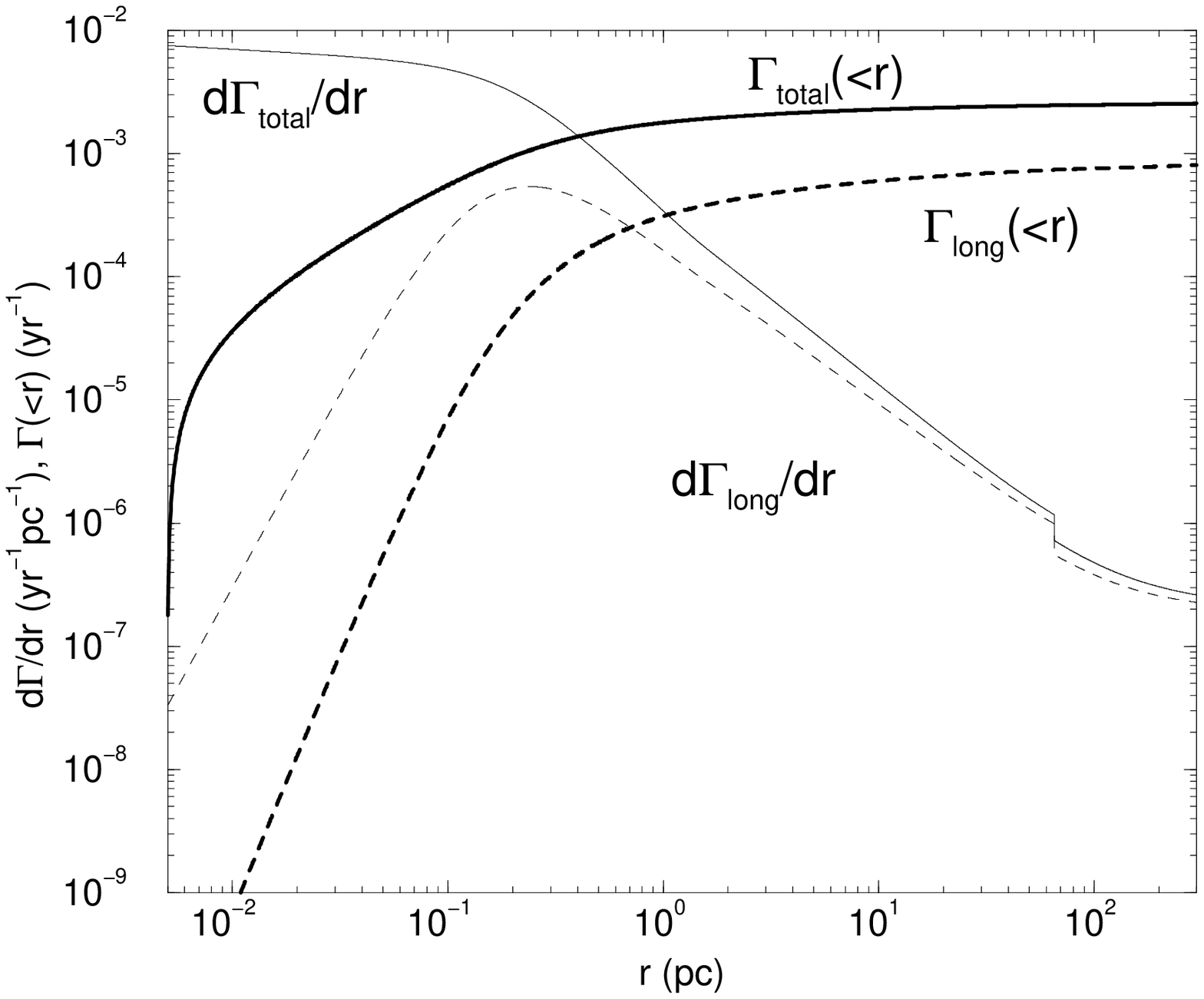}}
\centerline{Figure 3.}
\end{figure}

\clearpage
\begin{figure}
\centerline{\epsfxsize=6.0in\epsffile{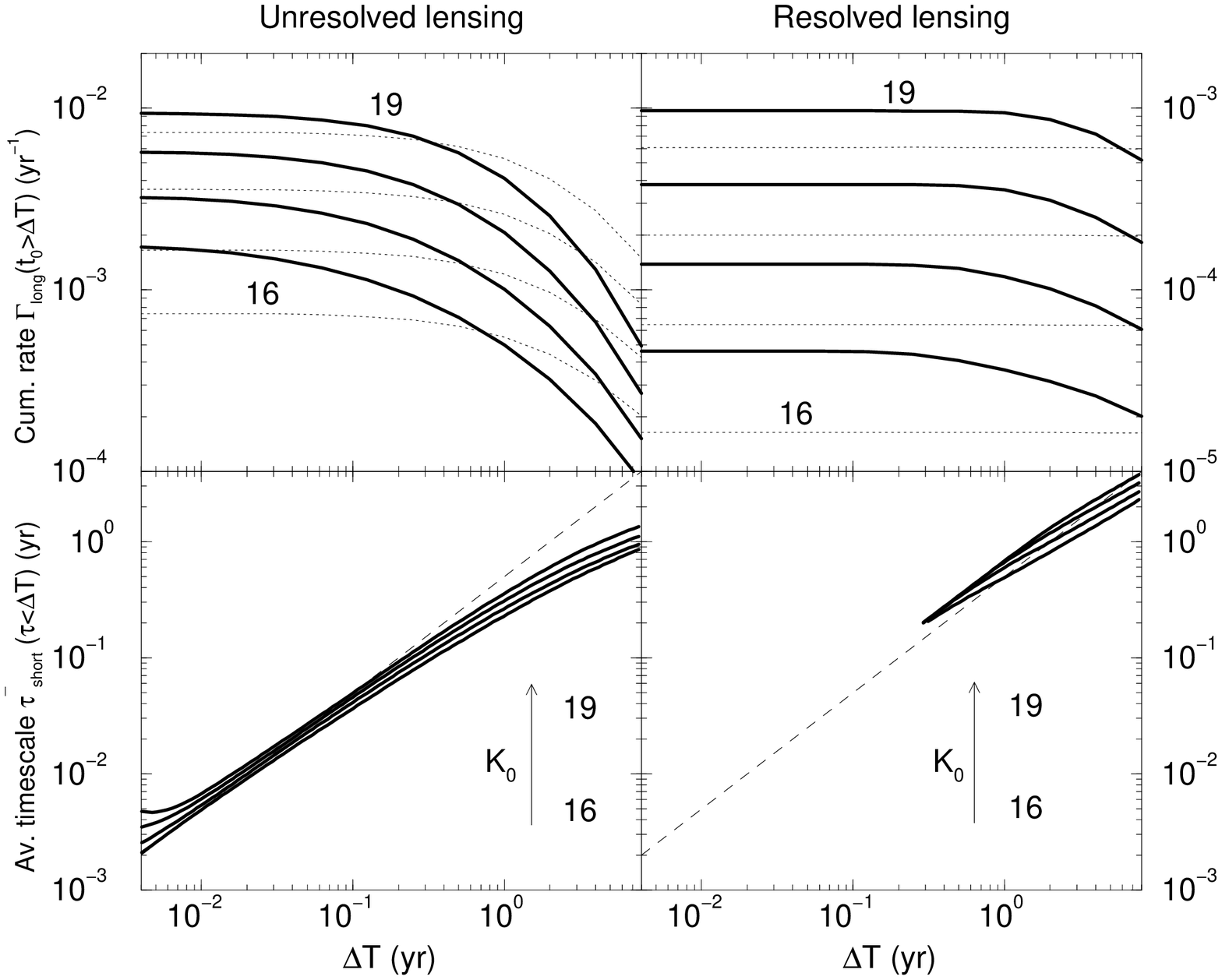}}
\centerline{Figure 4.}
\end{figure}

\clearpage
\begin{figure}
\centerline{\epsfxsize=6.0in\epsffile{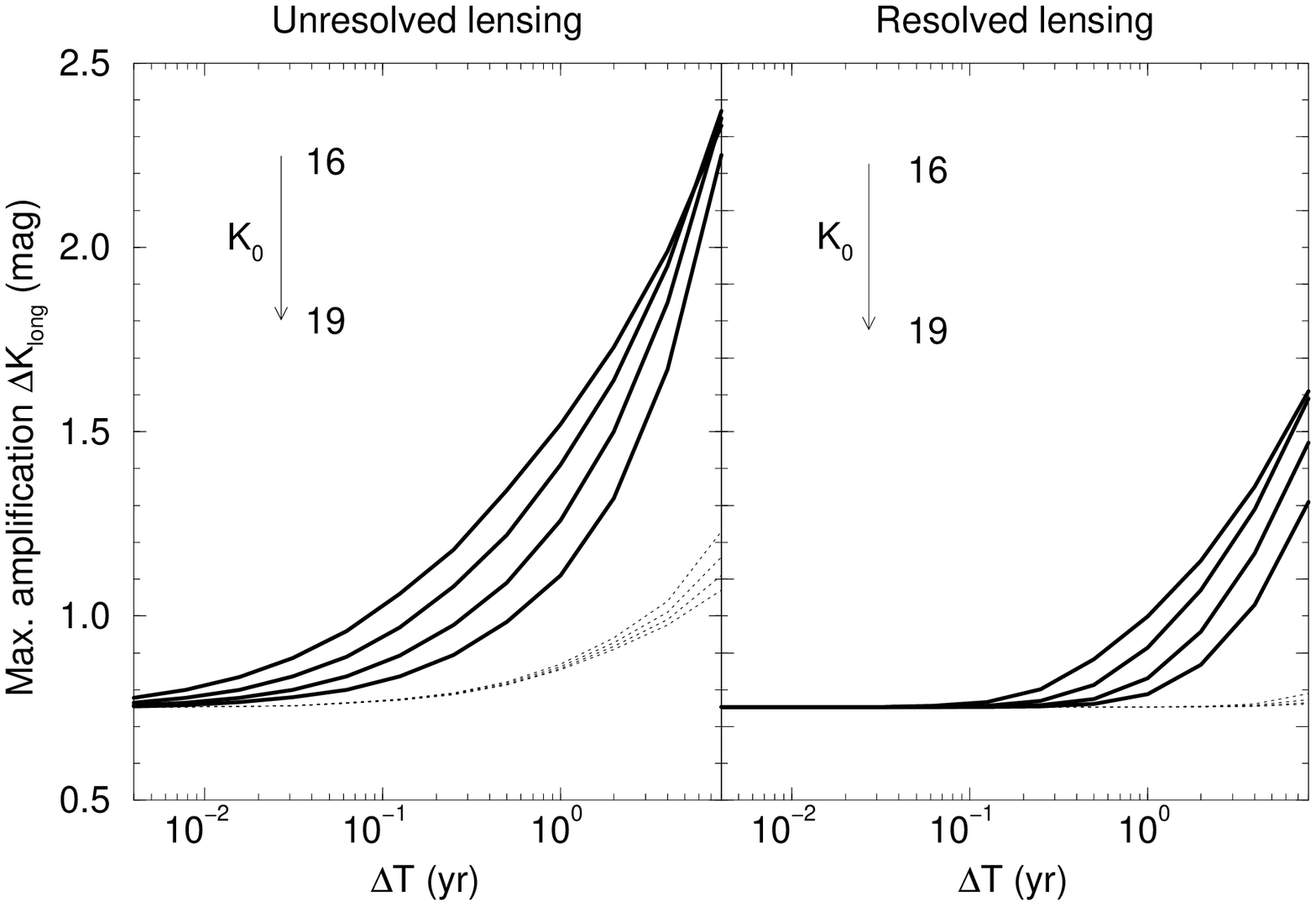}}
\centerline{Figure 5.}
\end{figure}

\end{document}